\documentclass[letterpaper]{jpconf}
\pdfoutput=1
\usepackage{graphicx}
\usepackage{maybemath}
\usepackage{hyperref}

\newcommand{\bdkstarmumu}{\maybebm{\mathrm{B}_\mathrm{d}\to\mathrm{K}^{*0}\mu^+\mu^-}}
\newcommand{\thetal}{\ensuremath{\theta_{\mathrm{L}}}}
\newcommand{\thetak}{\ensuremath{\theta_{\mathrm{K}}}}
\newcommand{\afb}{\ensuremath{A_{\mathrm{FB}}}}
\newcommand{\pT}{\ensuremath{p_{\mathrm{T}}}}

\begin{document}
\title{%
\maybebm{%
{\bdkstarmumu}} as a lab for discovering new physics at LHCb}

\author{Hugh Skottowe}

\address{Cavendish Laboratory, University of Cambridge}

\eads{\mailto{skottowe@hep.phy.cam.ac.uk}}

\vspace*{5pt}

\address{\textit{on behalf of the LHCb collaboration}}

\begin{abstract}
The analysis of the penguin decay
{\bdkstarmumu}
at LHCb can act as
a laboratory for the discovery and understanding of new physics.
Through the Operator Product Expansion, the decay kinematics are well
understood in both the Standard Model and in a large range of new
physics scenarios. The theoretical errors from QCD effects can be
characterized and a set of observables have been derived which
minimise their influence in measurements. We will describe how these
measurements can be made in LHCb with special emphasis on what can be
done with a first run of the LHC with a few hundred pb$^{-1}$ of
integrated luminosity.
\end{abstract}

%%%%%%%%%%%%%%%%%%%%%%%%%%%%%%%%%%%%%%%%%%%%%%%%%%%%%%%%%%%%%%%%%
%%%%%%%%%%%%%%%%%%%%%%%%%%%%%%%%%%%%%%%%%%%%%%%%%%%%%%%%%%%%%%%%%
\section{Introduction}
\begin{figure}
{\includegraphics[width=0.4\textwidth]{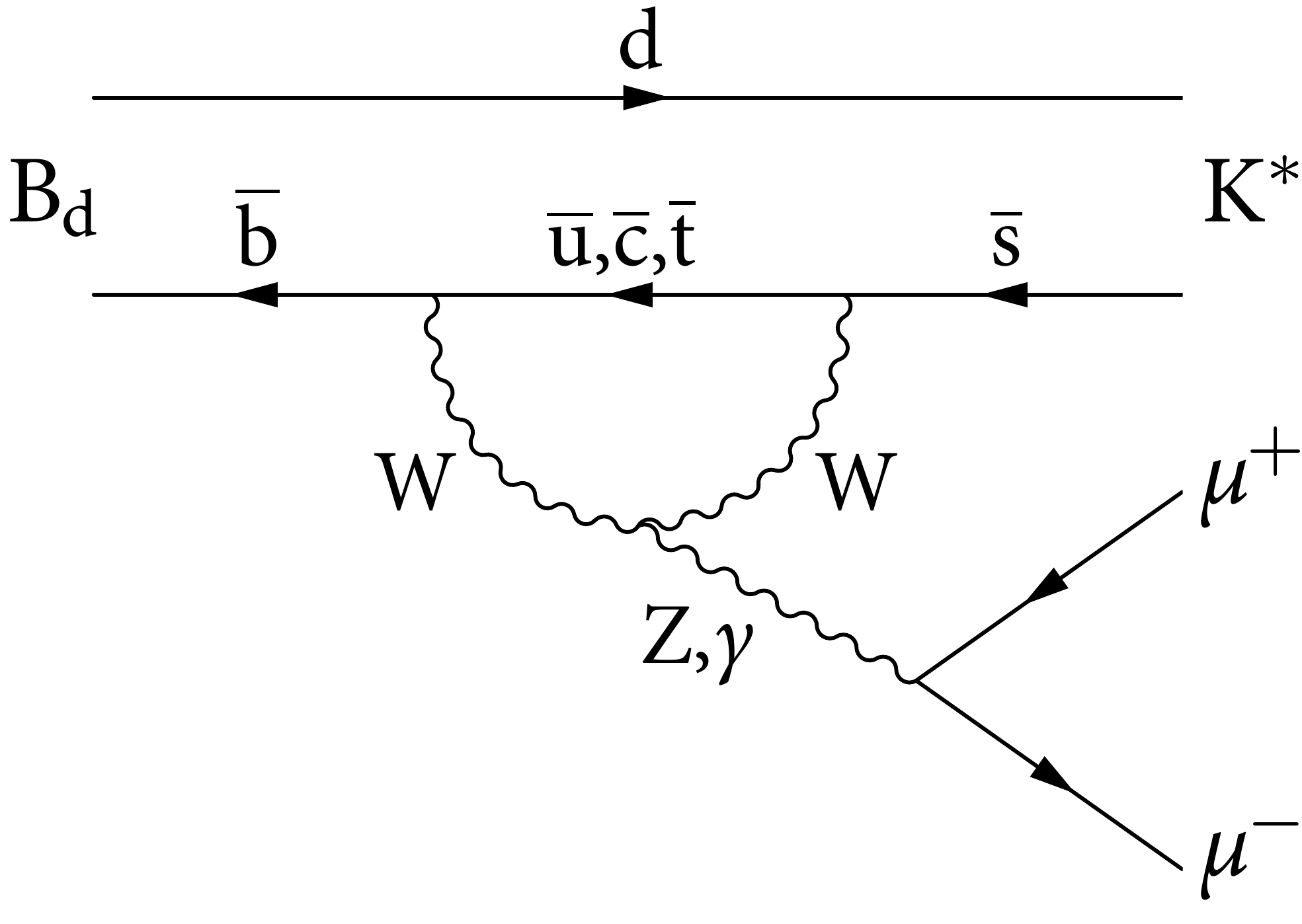}}
\hfill
{\includegraphics[width=0.4\textwidth]{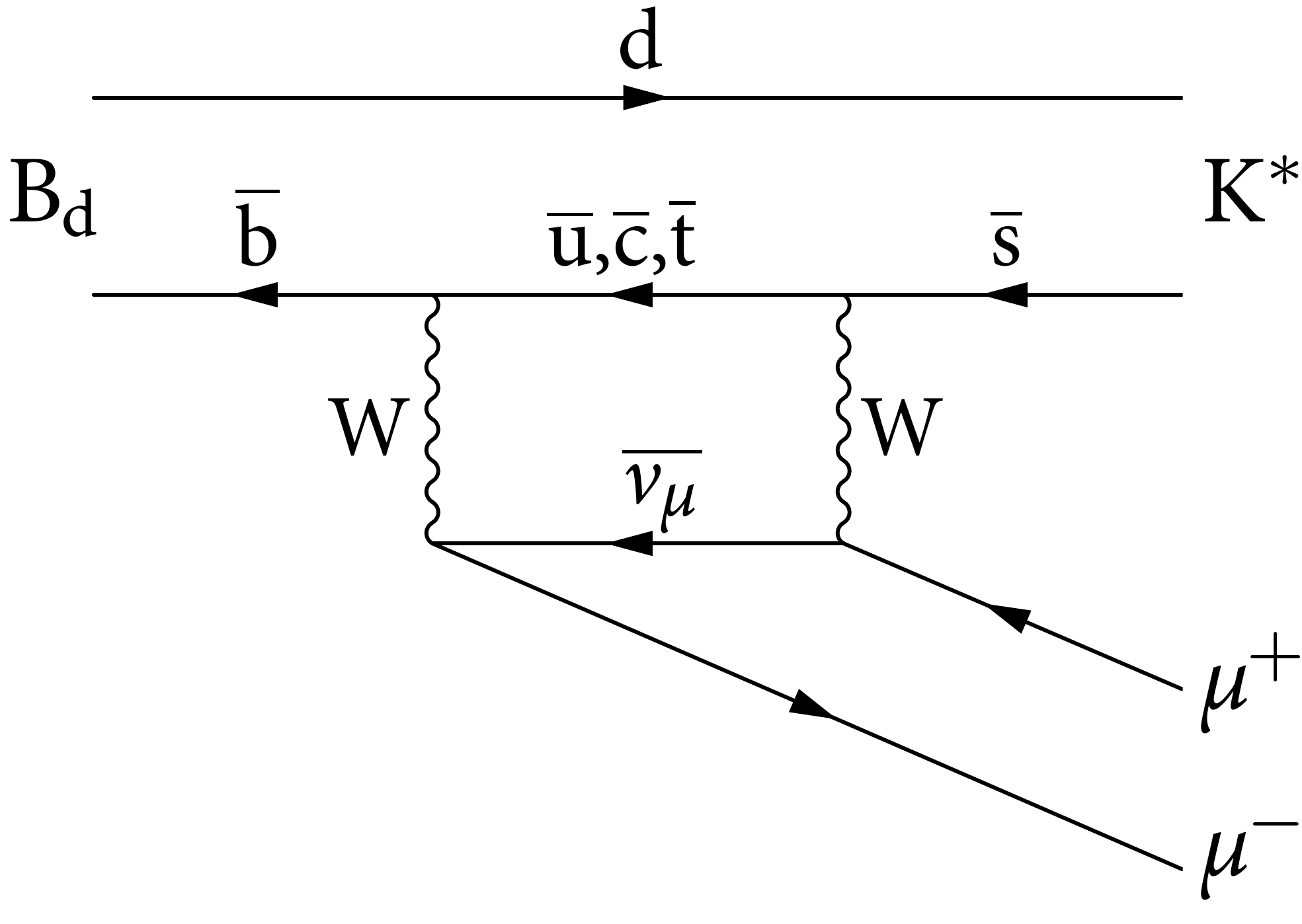}}
\caption{Example Feynman diagrams for the decay
{\bdkstarmumu} in the Standard Model.}\label{fig:feyn}
\end{figure}

As a flavour changing neutral current process,
{\bdkstarmumu} is suppressed in the Standard Model.
The decay occurs through a
$\overline{\mathrm{b}}\to\overline{\mathrm{s}}$ quark transition,
via a loop or box diagram, as shown in figure~\ref{fig:feyn}.

The measured branching ratio is
$(9.8\pm 2.1) \times 10^{-7}$~\cite{pdg}.
The kinematics of the decay can be fully described in terms of three angles, {\thetal}, {\thetak} and $\phi$, in addition to the di-muon invariant mass squared, $q^2$.
Many properties of the decay can be used as
powerful indirect searches for new physics.

The forward-backward asymmetry ({\afb}) of the muon pair
in this decay is one such interesting observable.
Formed from the lepton helicity angle {\thetal},
and varying as a function of $q^2$,
the hadronic uncertainties in the prediction of
this observable cancel when $\afb=0$,
and are smallest in the range $1<q^2<6\,\mathrm{GeV}^2\;\!\!/c^4$.

%%%%%%%%%%%%%%%%%%%%%%%%%%%%%%%%%%%%%%%%%%%%%%%%%%%%%%%%%%%%%%%%%
%%%%%%%%%%%%%%%%%%%%%%%%%%%%%%%%%%%%%%%%%%%%%%%%%%%%%%%%%%%%%%%%%
\section{Status of measurements of {\bdkstarmumu}}
The decay {\bdkstarmumu} has been observed at three independent experiments: the B-factories Belle and BaBar, and the CDF experiment at the Tevatron~\cite{otherexperiments}.
Each of these has recorded $\mathcal{O}(100)$ events, and measured
the branching ratio and $A_{\mathrm{FB}}(q^2)$.

%%%%%%%%%%%%%%%%%%%%%%%%%%%%%%%%%%%%%%%%%%%%%%%%%%%%%%%%%%%%%%%%%
%%%%%%%%%%%%%%%%%%%%%%%%%%%%%%%%%%%%%%%%%%%%%%%%%%%%%%%%%%%%%%%%%
\section{The LHCb experiment}
\begin{figure}\centering
\includegraphics[width=0.74\textwidth]{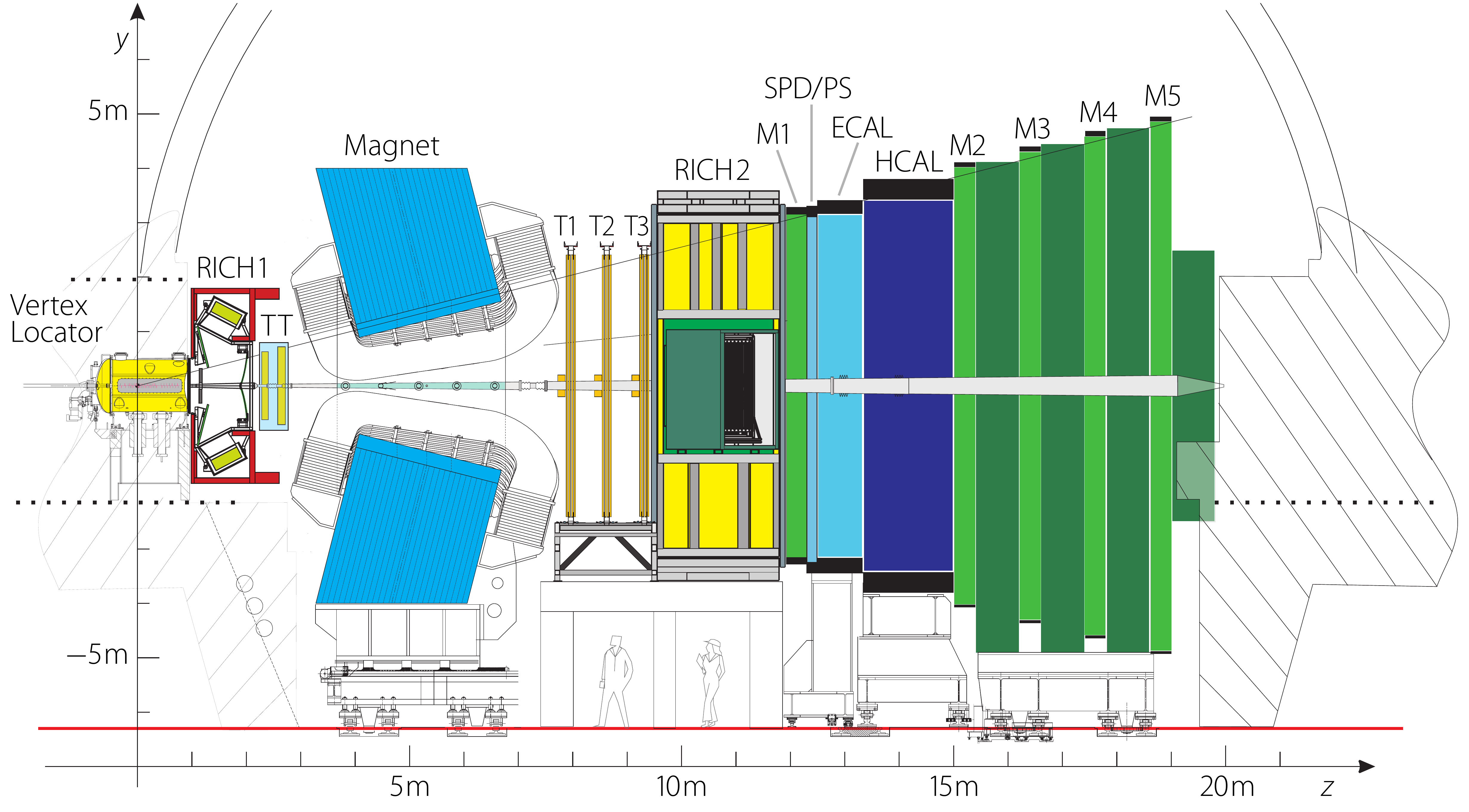}
\caption{LHCb, a forward spectrometer designed for b-physics measurements at the LHC.}\label{fig:detector}
\end{figure}

LHCb is a dedicated b-physics experiment at CERN's Large Hadron Collider (LHC).
The detector, shown in figure~\ref{fig:detector} and described
in detail in reference~\cite{detpaper}, has
a forward geometry, with an angular acceptance
from 10 to 250/300\,mrad.
There are several detector components, including a precise vertex locator, with silicon detector modules only 5\,mm from the proton beams.
In addition, two ring imaging Cherenkov detectors provide accurate particle identification, particularly for distinguishing between kaons and pions.
One of LHCb's priorities for the first LHC run, during 2010-2011,
is to measure {\afb} in {\bdkstarmumu}, 
where the 
K$^{*0}$ decays to K$^+\pi^-$.
The planned analysis is described in the following sections,
and in further detail in reference~\cite{roadmap}.

%%%%%%%%%%%%%%%%%%%%%%%%%%%%%%%%%%%%%%%%%%%%%%%%%%%%%%%%%%%%%%%%%
%%%%%%%%%%%%%%%%%%%%%%%%%%%%%%%%%%%%%%%%%%%%%%%%%%%%%%%%%%%%%%%%%
\section{Event trigger and selection}
The LHCb trigger is a two-level system.
First, a hardware-based `Level 0' trigger is used to reduce the 40\,MHz bunch crossing rate to 1\,MHz, using information from the muon and calorimetry detectors.
For the decay {\bdkstarmumu}, the event rate is reduced by applying a minimum value cut
to the transverse momentum of a single muon, or to the sum of the
transverse momentum of two muons

When an event is accepted by the Level 0 trigger, the full detector is read out
and the data passed to the software-based `High Level' trigger.
In the case of {\bdkstarmumu}, this trigger applies cuts to the
impact parameter and transverse momentum of a single muon, or 
to the impact parameter and vertex displacement of 
a muon and another charged track.

Once the triggered data is stored, it can then be analysed
offline. For the offline event selection, LHCb intends to use
a Fisher discriminant, applying a cut to separate signal
from background. Specific vetos are also used to remove
particle mis-identification backgrounds from 
$\mathrm{B}_{\mathrm{s}} \to \phi (\to\!\mathrm{KK}) \mu^+\mu^-$
and
$\mathrm{B}_{\mathrm{d}} \to X (\to\! Y \pi) \mathrm{J}/\psi (\to\!\mu^+\mu^-)$,
with generic $X$,$Y$.
Yields of $6200_{-1500}^{+1700}$ signal and $1550\pm 310$ background
events per 2\,fb$^{-1}$ are expected~\cite{fishernote}.

%%%%%%%%%%%%%%%%%%%%%%%%%%%%%%%%%%%%%%%%%%%%%%%%%%%%%%%%%%%%%%%%%
%%%%%%%%%%%%%%%%%%%%%%%%%%%%%%%%%%%%%%%%%%%%%%%%%%%%%%%%%%%%%%%%%
\section{Correcting the detector acceptance}
Before measuring observables such as {\afb},
the acceptance must be corrected.
The acceptance of events can vary as a function of the
lepton helicity angle, {\thetal},
and therefore shift the observed value of {\afb},
which measures the asymmetry of the {\thetal} distribution.
A non-flat acceptance in {\thetal} can be caused by the
detector and by the event trigger and selection processes.
Cuts on the {\pT} of both muons can greatly reduce the acceptance
 at low ($\sim$0) and high ($\sim${}$\pi$) values of {\thetal},
while reducing the acceptance much less at
intermediate {\thetal} values~\cite{roadmap}.
The LHCb detector has a similar, but less pronounced, effect,
because of the forward geometry
which only allows muons with
\mbox{$p>3\,\mathrm{GeV}\;\!\!\;\!\!/c$}
to reach the muon detectors.
This results in a non-flat acceptance because
decays with high or low values of {\thetal} have one muon
with high {\pT} and one with low {\pT}, whereas intermediate
values of {\thetal} are associated with muon pairs with
similar {\pT}.

The acceptance will be corrected by unfolding the acceptance that
is measured on simulated data samples of the signal decay.
Also under investigation is the use of a control channel,
$\mathrm{B}_{\mathrm{d}} \to \mathrm{K}^{*0} \mathrm{J}/\psi$,
to measure the acceptance.

%%%%%%%%%%%%%%%%%%%%%%%%%%%%%%%%%%%%%%%%%%%%%%%%%%%%%%%%%%%%%%%%%
%%%%%%%%%%%%%%%%%%%%%%%%%%%%%%%%%%%%%%%%%%%%%%%%%%%%%%%%%%%%%%%%%
\section{Measuring the forward-backward asymmetry}
Two methods have been considered by LHCb for measuring the
forward-backward asymmetry.
Firstly, accepted events can be binned in $q^2$ and in {\thetal}.
A value of {\afb} can then be calculated for each bin of $q^2$
from the numbers of forward and backward events in that $q^2$ bin.

Secondly, events can again be divided into two bins of {\thetal}, i.e.
into forward and backward categories.
For each of these two categories, the $q^2$ distribution of the events
is fitted with a third-order polynomial to find the functions
$N_{\textrm{forward}} (q^2)$ and $N_{\textrm{backward}} (q^2)$.
First-order polynomials, fitted to forward and backward
simulated backgrounds, are then subtracted.
The two fitted and subtracted polynomials,
representing forward and backward signal contributions
separately, can then be algebraically manipulated to find a functional form
of $A_{\mathrm{FB}} (q^2)$.

Both methods, the counting method using bins in both {\thetal} and $q^2$,
and the fitting of forward and backward events separately, achieve
approximately the same precision on a measurement of the zero
crossing point, $s_0^{\phantom{0}}$, (at which $\afb$=$0$)
in simulated data using the Standard Model prediction,
with a forecast sensitivity of
$\sigma (s_0^{\phantom{0}}) = 0.5\,\mathrm{GeV}^2\;\!\!/c^4$
with 2\,fb$^{-1}$ of
integrated luminosity.

%%%%%%%%%%%%%%%%%%%%%%%%%%%%%%%%%%%%%%%%%%%%%%%%%%%%%%%%%%%%%%%%%
%%%%%%%%%%%%%%%%%%%%%%%%%%%%%%%%%%%%%%%%%%%%%%%%%%%%%%%%%%%%%%%%%
\section{What can LHCb do with early data from the LHC?}

With data from the first LHC run, at 
$\sqrt{s}=7\,$TeV
during 2010-2011, LHCb intends to measure the forward-backward
asymmetry $A_{\mathrm{FB}}$, which has previously been measured
at BaBar, Belle and CDF.

\begin{figure}
{\includegraphics[width=0.44\textwidth]{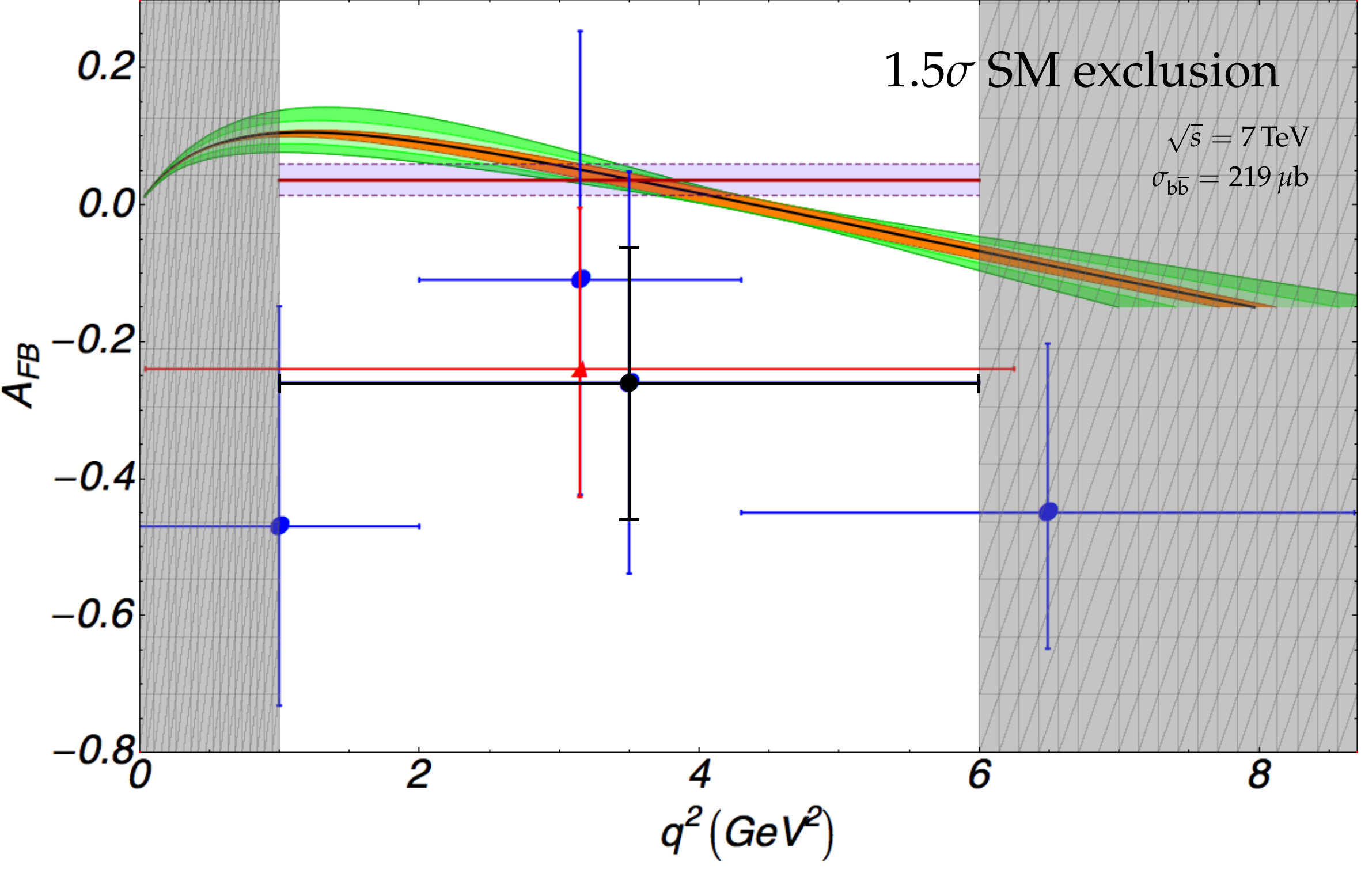}}
\hfill
{\includegraphics[width=0.44\textwidth]{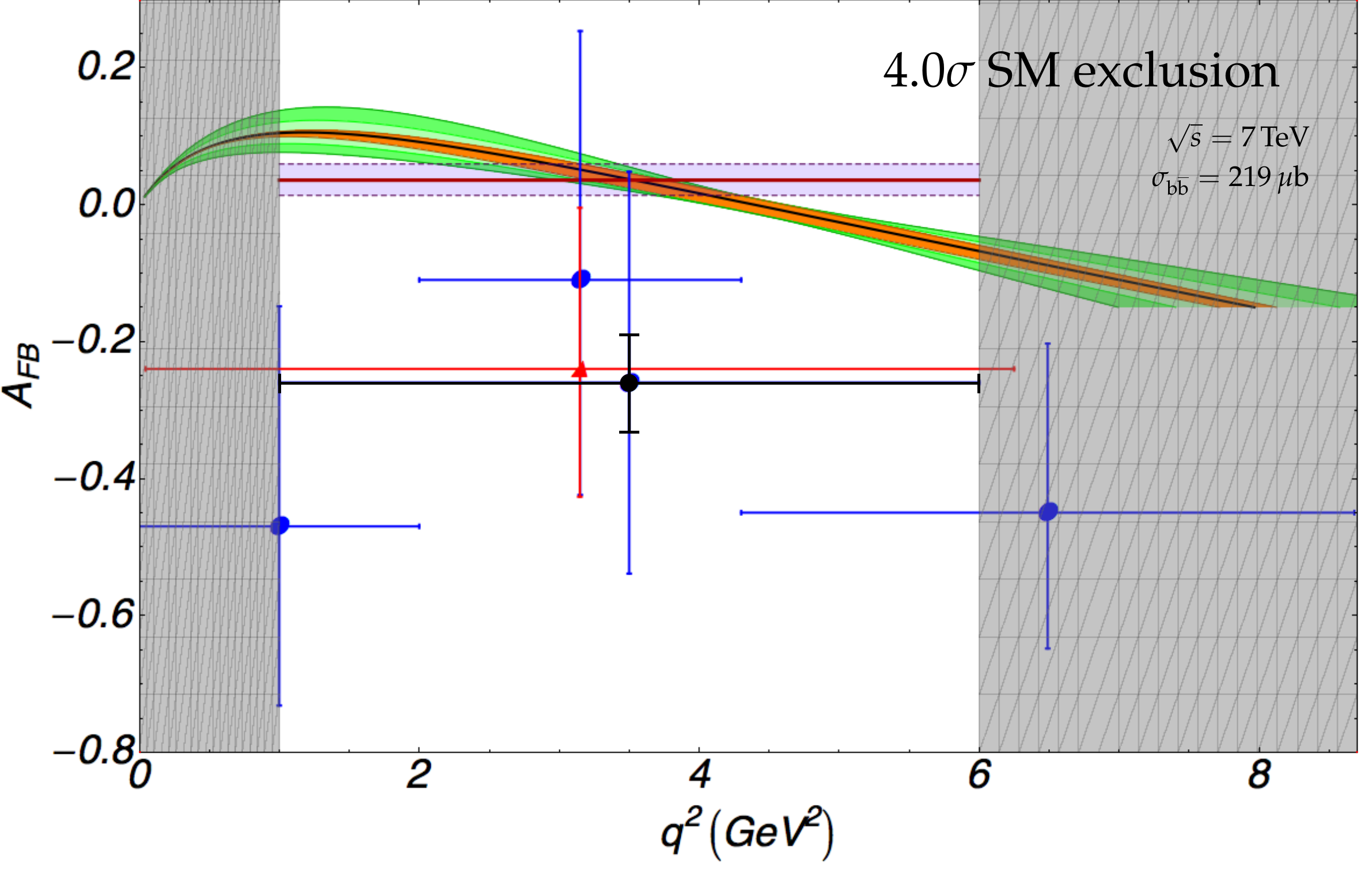}}

\vspace{-6pt}

{\small\hspace{36mm}(a)\hspace{84mm}(b)}

\vspace{-6pt}

\caption{Forecast sensitivity of LHCb
with (a) 0.1\,fb$^{-1}$ (b) 1\,fb$^{-1}$,
compared with published results from BaBar (red) and Belle (blue).
The Standard model prediction is
shown as a curve in orange and green, and as an average over
$1<q^2<6\,\mathrm{GeV}^2\;\!\!/c^4$ in pink~\cite{egedejhep2008}.
Forecast error bars for LHCb are shown in black on a data point that assumes
the same central value as the Belle measurement.
For this central value in an LHCb measurement, the Standard Model prediction
would be excluded by 1.5$\sigma$ after 0.1\,fb$^{-1}$, and 4.0$\sigma$
after 1\,fb$^{-1}$.}\label{fig:sens}
\end{figure}

Figure~\ref{fig:sens} shows predicted LHCb statistical precision for a
measurement of {\afb} in a single bin,
$1<q^2<6\,\mathrm{GeV}^2\;\!\!/c^4$,
for two different integrated luminosities, 0.1\,fb$^{-1}$ and 1\,fb$^{-1}$.
LHCb expects to select 1400 {\bdkstarmumu} events per fb$^{-1}$
for offline analysis.
The current LHC plan allows for 1\,fb$^{-1}$ of integrated proton-proton
luminosity to be delivered to the experiments during the 2010-2011 run.
As shown in the figure, LHCb will start to compete with
the B-factories BaBar and Belle with just 0.1\,fb$^{-1}$,
and will have significantly improved precision with 1\,fb$^{-1}$.

%%%%%%%%%%%%%%%%%%%%%%%%%%%%%%%%%%%%%%%%%%%%%%%%%%%%%%%%%%%%%%%%%
%%%%%%%%%%%%%%%%%%%%%%%%%%%%%%%%%%%%%%%%%%%%%%%%%%%%%%%%%%%%%%%%%
\section{What can LHCb do with more data?}

After the 2010-2011 run, an LHC shutdown is planned,
followed by further running, 
up to the design energy, 
$\sqrt{s}=14\,$TeV.
Integrated luminosity will rapidly increase beyond
the 1\,fb$^{-1}$ that is expected during 2010-2011.
With the larger dataset, LHCb will continue to
improve the precision on its measurement of $A_{\mathrm{FB}}$.
In addition, the increased data will allow several
further measurements of the decay.
Firstly, projections of the three angles {\thetal}, {\thetak} and $\phi$
will become possible.
This will allow the measurement of the longitudinal polarization, $F_{\mathrm{L}}$,
for which very precise theoretical predictions can be made.

With more data, it will become possible to perform a full angular
fit.
Fitting the three angles {\thetal}, {\thetak} and $\phi$, simultaneously with
$q^2$, will enable the measurement of a wide range of further
observables with sensitivity to new physics.
For example, the $A_{\mathrm{T}}^{(2)}$ observable is very sensitive
to the right-handed $\mathcal{C}_7'$ Wilson coefficient~\cite{at2},
and the $S_5$ observable is sensitive to a range of supersymmetric
models~\cite{s5}.

%%%%%%%%%%%%%%%%%%%%%%%%%%%%%%%%%%%%%%%%%%%%%%%%%%%%%%%%%%%%%%%%%
%%%%%%%%%%%%%%%%%%%%%%%%%%%%%%%%%%%%%%%%%%%%%%%%%%%%%%%%%%%%%%%%%
\section{Conclusions}
The decay {\bdkstarmumu} has properties that
are precisely predicted both in the Standard Model, 
and in various new physics models.
LHCb, a dedicated b-physics experiment at the LHC,
is ideally suited to studying this decay, using it
as an indirect probe of new physics.

LHCb will quickly collect large numbers of {\bdkstarmumu}
events.
Signal (background) yields of 6200$_{-1500}^{+1700}$
($1550\pm 310$) events are expected per nominal year of 2\,fb$^{-1}$
with %7+7\,TeV
$\sqrt{s}=14\,$TeV,
assuming a b$\overline{\mathrm{b}}$ production cross-section of
$\sigma_{\mathrm{b}\overline{\mathrm{b}}} = 500\,\mu$b.
At the currently planned energy of 
$\sqrt{s}=7$\,TeV,
yields of approximately 1400 signal events
are expected during a
\mbox{2010-11}
run of $\sim$1\,fb$^{-1}$.
This number of events is sufficient for LHCb's precision
to compete with previous results during this first data run.
LHCb will then go on to make first measurements of further observables
that are powerful probes of new physics.

%\section{Introduction}
%The basic style file used is that of  \jpcs. All proceedings must be produced using \LaTeX\ and must also be submitted to the arXiv server where they will be linked from a conference page.

%\section{Preparing your paper}
%This document uses the basic features of the style file
%but a full description of all the features can be found in docs directory under "JPCSLaTeXGuidelines.pdf". The associated \LaTeX\ file also uses the style file. Please note that the page size should be set to {\em letter and not A4} because, despite Canada being metric, we can only purchase paper in US sizes. To do this ensure that your document starts with:

%\begin{verbatim}
%\documentclass[letterpaper]{jpconf}
%\end{verbatim}

%A Bibtex style, "iopart-num", file is also included. Documentation on how to use this is given in the file "bibtex-style.pdf" should you wish to use Bibtex.

%\section{Refereeing}
%Please note that your document will not be refereed but will be proof-read for typos, spelling and grammar. To ensure consistency Canadian english spellings should be used for all words e.g. colour, centre, metre, flavour etc. After proof-reading a list of corrections will be sent to you. However, since we cannot update your arXiv paper directly it will be up to you to implement these corrections in the online version (this may change in future years). The edited version will be the one appearing in the printed proceedings though.

\end{document}